\documentclass[%
 aip,
 amsmath,amssymb,
 reprint,%
]{revtex4-1}

\usepackage{graphicx}
\usepackage{dcolumn}
\usepackage{bm}

\usepackage[utf8]{inputenc}
\usepackage[T1]{fontenc}
\usepackage{mathptmx}
\usepackage{etoolbox}
\usepackage{xcolor}
\usepackage{ulem}

\newtheorem{thm}{Theorem}

\newtheorem{lemma}[thm]{Lemma}

\newtheorem{remark}[thm]{Remark}

\newcommand{\abs}[1]{\left\vert#1\right\vert}
\makeatletter
\def\@email#1#2{%
 \endgroup
 \patchcmd{\titleblock@produce}
  {\frontmatter@RRAPformat}
  {\frontmatter@RRAPformat{\produce@RRAP{*#1\href{mailto:#2}{#2}}}\frontmatter@RRAPformat}
  {}{}
}%
\makeatother
\begin{document}

\preprint{AIP/123-QED}

\title[Expression of the peak time for time-domain boundary measurements in diffuse light]{Expression of the peak time for time-domain boundary measurements in diffuse light}
\author{J.Y. Eom}
\affiliation{Graduate School of Mathematical Sciences, The University of Tokyo,  
Tokyo 153-8914, Japan}%

\author{M. Machida}%
\affiliation{Institute for Medical Photonics Research, Hamamatsu University School of Medicine, \\
Hamamatsu 431-3192, Japan}%

\author{G. Nakamura}%
\affiliation{Department of Mathematics, Hokkaido University, Sapporo 060-0810, Japan and Research Center of Mathematics for Social Creativity, Research Institute for Electronic Science, Hokkaido University, Sapporo, 060-0811, Japan}%

\author{G. Nishimura}%
\affiliation{Research Institute for Electronic Science, Hokkaido University, Sapporo 001-0020, Japan
}%

\author{C.L. Sun}
\affiliation{College of Science, Nanjing University of Aeronautics and Astronautics,
Nanjing 211106, P.R. China}%
\affiliation{Nanjing Center for Applied Mathematics, Nanjing 211135, P.R.China.}%

\email{sunchunlong@nuaa.edu.cn}

\date{\today}

\begin{abstract}
Light propagation through diffusive media can be described by the diffusion equation in a space-time domain. Further, fluorescence can be described by a system of coupled diffusion equations. This paper analyzes time-domain  measurements, which measure the temporal point-spread function (TPSF), at a boundary of such diffusive media with a given source and detector. We focus on the temporal position of the TPSF maximum, which we refer to as the peak time.  Although some unique properties of solutions of this system have been numerically studied, we give a mathematical analysis of peak time, providing proof of the existence, uniqueness, and the explicit expression of the peak time. We clearly show the relationship between the peak time and the object position in a medium. 
\end{abstract}

\maketitle


\section{Introduction}

Light propagation in highly scattering medium, such as biological tissue, is dominated by multiple scattering. The propagation can be described by an initial-boundary value problem for a diffusion equation in a space-time  domain given as
\begin{eqnarray}\label{excitation-Ue}
\begin{cases}
\left( \frac{1}{c} \frac{\partial}{\partial t} -\nabla \cdot D\nabla +\mu_a\right)u_e=0, \quad (x,t)\in\Omega_T:=\Omega\times (0,T),\\
\nu\cdot\nabla u_e+\beta u_e=\delta(x-x_s)\delta(t), \quad x\in\partial\Omega, \; t\in(0,T),\\
u_e(x,0)=0, \quad x\in\Omega,
\end{cases}
\end{eqnarray}
where $\Omega\subset\mathbb{R}^3$ represents the specified background medium with boundary $\partial\Omega$, $u_e(x,t)$ is the photon density of light, $\nu$ is the unit outward normal, $c$ is the speed of light in the medium, $D=D(x)>0$ is the photon diffusion coefficient and $\mu_a=\mu_a(x)>0$ is the absorption coefficient of medium\cite{Marttelli,Jiang11, Durduran10}. Here $x_s\in\partial\Omega$ is the location of point source and $\beta$ is a parameter given by
\begin{equation*}
\beta=\frac{1}{2D} \frac{1-2\int_0^1 \mathcal R(\mu)\mu \, {\rm d}\mu}{1+3\int_0^1 \mathcal R(\mu)\mu^2 \, {\rm d}\mu}
\end{equation*}
with the Fresnel reflectance $\mathcal R(\mu)$, which depends on the refractive index ratio between the medium and the free space \cite{Marttelli}. 

The initial-boundary value problem \eqref{excitation-Ue} is usually used for the light propagation in highly scattering media, such as biological tissues  and is applied for the quantitative analysis of the optical properties of the medium\cite{ Mycek03, Rudin13, Nitziachristos V02, Arridge99, Arr09}. This kind of analysis is so-called diffuse optical spectroscopy (DOS) or more sophisticated one, diffuse optical tomography (DOT). The DOS or DOT is to identify the unknown absorption coefficient $\mu_a(x)$ and diffusion coefficient  $D(x)$ from the time-domain (TD) measurements at the boundary
\begin{equation}\label{data-ue}
u_e(x,t;x_s),\quad x,\; x_s\in\partial\Omega,\;t\in(0,T),
\end{equation}
which depend on both of the space and time.

On the other hand, fluorescence in highly scattering medium is also very important in the applications and can be quantified by a similar initial-boundary value problem.  In case of the fluorescence, two physical processes are coupled, namely excitation and fluorescence (emission). First, the excitation light injected from the boundary propagate in the medium as \eqref{excitation-Ue} and then absorbed by the fluorophores. Next, fluorescence emitted from the fluorophores propagate in the medium and is finally detected by a detector on the boundary described by  
\begin{eqnarray}\label{emission-Um}
\begin{cases}
\left( \frac{1}{c} \frac{\partial}{\partial t} -\nabla \cdot D\nabla +\mu_a \right) U_m= S[\mu_f,u_e](x, t;\tau), \quad (x,t)\in\Omega_T,\\
\nu\cdot\nabla U_m+\beta U_m=0, \quad (x,t)\in\partial\Omega\times(0,T),\\
U_m(x,0)=0, \quad x\in\Omega.
\end{cases}
\end{eqnarray}
In general, $D$ and $\mu_a$ here are not the same as the excitation ones in \eqref{excitation-Ue}. We assume these are the same in this paper, corresponding to a case when the excitation and emission wavelengths are close and $\mu_f\ll\mu_a$. 
The source term $S$ for $U_m$ on the right-hand side of \eqref{emission-Um} contains the excitation field $u_e$ and is specified by
\begin{equation}\label{source term}
S[\mu_f,u_e](x,t;\tau)=\frac{\mu_f(x)}{\tau}\int_0^t e^{-\frac{t-s}{\tau}} u_e(x, s; x_s)\,{\rm d}s,
\end{equation}
where $\tau > 0$ is the fluorescence lifetime and $\mu_f(x)>0$ is the absorption coefficient of fluorophores. Here we assume the optical parameters for fluorescence are same to those for the excitation and the absorption of the fluorophores is sufficiently smaller than the absorption of the medium. The inverse problem to identify the unknown absorption coefficient $\mu_f(x)$ from the TD boundary measurements
\begin{equation}\label{data-um}
U_m(x,t;x_s),\quad x,\; x_s\in\partial\Omega, \;t\in(0,T)
\end{equation}
is referred to as the TD fluorescence diffuse optical tomography (FDOT)\cite{Lam05, Han08, Gao08}.

For above inverse problems of DOS and FDOT, other two types of the measurements, continuous wave (CW) \cite{Patwardhan05, Ducros11} and frequency domain (FD) \cite{Nitziachristos01, Corlu-etal07, Milstein04}, are also used. Among above three types of measurements, the TD measurements measure the temporal point-spread function (TPSF) and will provide the most fruitful information. Even though this advantage, it is believed that the effective choice of TPSF data is needed to extract important  information about the unknown target. The possible choices of the TPSF data and the relative robustness of the different sets of data to noise are considered \cite{J.Riley07}. They refer the data in the whole TPSF  as global data types, whereas the types of information only local to the high signal area are referred to local data types. It was examined that the local data types are more robust to noise than global data types, and should provide enhanced information to the related inverse problems. 

In this paper, we denote the temporal position of the maximum of TPSF as peak time $t_{\rm peak}$ and assume $t_{\rm peak}<T$. One has demonstrated direct depth estimation of a localized fluorescent object from the peak time by numerical experiments\cite{D. Hall04}. The depth estimation permits recovery of the fluorophore concentration, which is a robust and efficient approach since the peak time is independent of the fluorophore concentration. However, the rigorous mathematical analysis of the peak time has not been provided yet as far as we know. In case of $u_e$, one can find an explicit expression of the peak time of TD boundary measurements \eqref{data-ue} but for the Direchlet boundary condition case ($\beta\rightarrow \infty$)\cite{G.Nishi05}. We will give the explicit expressions of peak time for the TD boundary measurements \eqref{data-ue} and \eqref{data-um}, by which we can construct the connection between the peak time and unknown important information in the related inverse problems. 

A mathematically well known hot spots problem is to extract which spatial point is the hottest point \cite{Siudeja15,Rodrigo99,Burdzy99}, whereas the peak time problem in this paper is to extract the time point at which the photon density of TD boundary measurements is strongest. Both of them are the problems of studying the evolution of solution's maximum value. Searching general schemes to identify a profile with peaks for convolutions of functions has been studied in signal processing \cite{Vidmar03,Shensa92,lin06}. For instance, there is a scheme using continuous wavelet transforms\cite{lin06}. We tried in this paper to derive the peak time by using an asymptotic analysis.

The rest of this paper is organized as follows. In section \ref{sec2}, under some simple assumptions, we show the asymptotic behavior of TD boundary measurements. Section \ref{sec3} shows the explicit expressions of the peak time for the boundary excitation measurements \eqref{data-ue} and the boundary emission measurements \eqref{data-um}, respectively. The unique existence of peak time is also considered in this section. Finally, section \ref{sec4} is devoted to conclusion and remark. 

\section{Asymptotic behavior of the boundary measurements} \label{sec2}
In the measurement setup of DOT or FDOT, the source and detection optical fibers are placed on the boundary surface $\partial\Omega$. We are here considering measurements, which are a set of source and detector pair on the same surface of biological tissue, and the distance between source and detector is very small rather than the tissue size.  In this case, we can assume the half space for the modeling of tissue. Hence we can model $\Omega$ as 
\begin{equation*}
\Omega:=\mathbb{R}^3_+=\left\{x=(x_1,x_2,x_3): (x_1,x_2)\in\mathbb{R}^2, x_3>0\right\}
\end{equation*}
with the boundary $\partial\Omega:=\left\{x=(x_1,x_2,0): (x_1,x_2)\in\mathbb{R}^2 \right\}$. On the other hand, we assume that the absorption coefficient $\mu_a$ and the diffusion coefficient $D$ are constants everywhere in the medium. In this paper we consider the zero-lifetime case of model \eqref{emission-Um}, i.e., the lifetime in the source term $S[\mu_f,u_e](x, t;\tau)$ is $\tau=0$. Furthermore, we assume the size of fluorescence target in \eqref{emission-Um} is very small such that it can be approximated by
\begin{equation}\label{delta}
\mu_f(x)=\delta(x-x_c),
\end{equation}
where $x_c=(x_{c1},x_{c2},x_{c3})\in\Omega$ is the location of point target.

Under above assumptions, we are going to obtain the analytical expressions of excitation light and zero-lifetime emission light. Let $K(x,y;t)$ be the Green function, which satisfies
\begin{eqnarray*}\label{Green function}
\begin{cases}
\left( \frac{1}{c} \frac{\partial}{\partial t} - D\Delta +\mu_a \right) K=\delta(x-y)\delta(t), \quad (x, t)\in\Omega_T,\\
K=0, \quad x\in\Omega, \; t=0,\\
\nu\cdot\nabla K + \beta K= 0, \quad (x, t)\in\partial\Omega\times (0,T)
\end{cases}
\end{eqnarray*}
for a given point source located at $y\in\Omega$. One can obtain the analytical expression of $K(x,y;t)$ by the functional solution of heat equation \cite{Haskell}, which has been investigated extensively in the scientific community. That is
\begin{equation}\label{expression of K}
K(x, y; t) = \frac{ce^{-c\mu_a t}}{(4\pi cD t)^{3/2}}
e^{-\frac{(x_1-y_1)^2+(x_2-y_2)^2}{4cDt}} K_3(x_3,y_3;t),
\end{equation}
where 
\begin{eqnarray*}
K_3(x_3,y_3;t) &=& e^{-\frac{{(x_3+y_3)}^2}{4cDt}} +e^{-\frac{{(x_3-y_3)}^2}{4cDt}} \\
&& - 2\beta\sqrt{\pi cDt}e^{\beta (x_3+y_3)+\beta^2cDt}
\mathop{\mathrm{erfc}}
\left(\frac{x_3+y_3+2\beta cDt}{\sqrt{4cDt}}\right)
\end{eqnarray*}
with the complementary error function 
\begin{equation*}
{\rm erfc}(\xi)=\frac{2}{\sqrt{\pi}}\int_\xi^\infty e^{-s^2}\,{\rm d}s,\,\xi\in\mathbb{R}.
\end{equation*}
Then, by the general theory of partial differential equations, we have the following expression for $u_e$ as
\begin{equation}\label{expression of u_e}
u_e(x,t;x_s)= D\times K(x,x_s;t), \quad (x,t)\in \Omega\times(0,T).
\end{equation}

Next, we consider the analytical solution of zero-lifetime emission light. Let $u_m$ be the photon density of zero-lifetime emission light. For $t\in(0,T)$, integrating \eqref{source term} by parts with respect to $s$ gives
\begin{eqnarray*}
S[\mu_f,u_e](x,t;\tau)&=&\frac{\mu_f(x)}{\tau}\int_0^t e^{-(t-s)/\tau} u_e(x,s;x_s)\,{\rm d}s\\
&=&\mu_f(x) \left\{ u_e(x,t;x_s) - u_e(x,0;x_s) e^{-\frac{t}{\tau}}  \right.\\
&&  \left.- \int_0^t \frac{\partial u_e(x,s;x_s)}{\partial s} e^{-\frac{t-s}{\tau}} \, {\rm d}s \right\}.
\end{eqnarray*}
Since $x_s\in\partial\Omega$, 
\begin{equation*}\label{smoothness of u_e}
u_e(x,t;x_s)\in C^\infty(\Omega\times[0,T])
\end{equation*}
due to \eqref{expression of u_e}. This implies that for each fixed $x\in\Omega$, $\frac{\partial u_e(x,s;x_s)}{\partial s}$ is bounded with respect to $s\in (0,T)$. Hence there exists a constant $M_x>0$ depending on $x$ such that
\begin{eqnarray*}
\abs{ \int_0^t \frac{\partial u_e(x,s;x_s)}{\partial s} e^{-\frac{t-s}{\tau}} \, {\rm d}s}
\leq M_x\int_0^t{e^{-\frac{t-s}{\tau}}} \, {\rm d}s = M\tau(1-e^{-\frac{t}{\tau}})
\end{eqnarray*}
for $t\in (0,T)$.
Together with this and $\lim_{\tau\to {0+}}{e^{-\frac{t}{\tau}}}=0$, we immediately have for $t>0$ that
\begin{eqnarray*}
\lim_{\tau\to 0+} S[\mu_f,u_e](x,t;\tau)=\mu_f(x) u_e(x,t;x_s),\,\,(x,t)\in\Omega_T.
\end{eqnarray*}
Hence the photon density of zero-lifetime emission light $u_m$ satisfies the following initial-boundary value problem
\begin{eqnarray}\label{sun2-4}
\begin{cases}
\left(\frac{1}{c} \partial_t - D\Delta +\mu_a\right) u_m = \mu_f(x)u_e(x,t;x_s),\quad (x,t)\in\Omega_T,\\
u_m(x,0)=0, \quad x\in\Omega,\\
\nu\cdot\nabla u_m + \beta u_m = 0, \quad (x,t)\in\partial\Omega\times (0,T),
\end{cases}
\end{eqnarray}
which implies the solution $U_m(x,t;x_s)$ to initial-boundary value problem \eqref{emission-Um} is the convolution of zero-lifetime emission $u_m(x,t;x_s)$ with the lifetime function $\frac{1}{\tau}e^{-t/\tau}, \tau>0$, i.e.,
\begin{equation*}
U_m(x,t; x_s)=\int_0^t \frac{1}{\tau}e^{-s/\tau} \, u_m(x, t-s; x_s) \, {\rm d}s, \; (x,t)\in{\overline\Omega}\times [0,T].
\end{equation*}

Now, by using the Green function and the expression of $u_e$ as in \eqref{expression of u_e}, we have the following analytical expression for $u_m$ as
\begin{equation}\label{um-point}
u_m(x,t;x_s)=D \int_0^t K(x,x_c;t-s) K(x_c,x_s;s) \,{\rm ds}.
\end{equation}
In particular, for any given $x, x_s\in\partial\Omega$, we have by \eqref{expression of K}, \eqref{expression of u_e} and \eqref{um-point} that
\begin{equation}\label{ue-mesure}
u_e(t) = \frac{ e^{-c\mu_a t}}{4\pi^{3/2} \sqrt{Dc}} t^{-\frac{3}{2}} e^{-\frac{\|x-x_s\|^2}{4cDt}} {\widetilde K}_3(0,0;t),
\end{equation}
 and
\begin{eqnarray}\label{mesure}
u_m(t) &=& \frac{ e^{-c\mu_a t}}{16\pi^3 D^2c} \int_0^t \frac{1}{\big[(t-s)s\big]^{3/2}}  e^{-\frac{\|x-x_c\|^2}{4cD(t-s)}} e^{-\frac{\|x_s-x_c\|^2}{4cDs}} \nonumber \\
&&\times {\widetilde K}_3(0,x_{c3};t-s) {\widetilde K}_3(x_{c3},0;s) \,{\rm d}s,
\end{eqnarray}
where
\begin{eqnarray*}
{\widetilde K}_3(x_3,y_3;t) &:=& 
1 - \beta\sqrt{\pi cDt} \,\exp{\left(\left(\frac{x_3+y_3+2\beta cDt}{\sqrt{4cDt}}\right)^2\right)} \nonumber \\
&&\times \mathop{\mathrm{erfc}}
\left(\frac{x_3+y_3+2\beta cDt}{\sqrt{4cDt}}\right).
\end{eqnarray*}

In what follows, if there is no specification, we will define
\begin{equation}\label{parameter}
C(t):=\frac{ e^{-c\mu_a t}}{16\pi^3 D^2c}, \;\;A:=\frac{\|x-x_c\|^2}{4cD}, \;\; B:=\frac{\|x_s-x_c\|^2}{4cD},
\end{equation}
where $\Vert\xi\Vert$ is the Euclidean norm of any three dimensional vector $\xi$. Then, by \eqref{parameter}, we can write $u_m$ of \eqref{mesure} as
\begin{equation}\label{eq:3.1}
u_m(t) = C(t) \int_0^t f_{A,\beta}(t-s)f_{B,\beta}(s)\; {\rm d}s,\quad t>0,
\end{equation}
where 
\begin{equation}
\label{eq:3.2}
f_{A,\beta}(s) := s^{-\frac{3}{2}}e^{-\frac{A}{s}} H_\beta(s),\quad
f_{B,\beta}(s) := s^{-\frac{3}{2}}e^{-\frac{B}{s}}H_\beta(s),
\end{equation}
with
\begin{equation}
\label{eq:3.3}
H_\beta(s) := 1 - \beta\sqrt{\pi cDs} e^{\xi^2} \mathrm{erfc}(\xi), \;\;
\xi:=\frac{x_{c3}+2\beta cDs}{\sqrt{4cDs}}>0
\end{equation}
for $0<s<t$. We will consider the peak time for \eqref{eq:3.1}. More precisely, we will consider the explicit expressions of approximate peak time for \eqref{eq:3.1} in three cases where $\beta$ in the boundary condition is really small, large and general value, respectively. This implies that we need to construct the connections between \eqref{eq:3.1} with the TD boundary measurements corresponding to Neumann boundary condition and Dirichlet  boundary  condition, respectively. We do that by analyzing the asymptotic behavior of TD boundary measurements for $\beta$ as follows.

\begin{lemma}\label{thm-asym}
Corresponding to the case of $\beta=0$, we set
\begin{equation}\label{tid-um}
\widetilde{u}_m(t) = C(t) \int_0^t f_{A,0}(t-s) f_{B,0}(s)\; {\rm d}s,
\end{equation}
which is the boundary measurements of \eqref{sun2-4} corresponding to Neumann boundary condition. Then, there holds the following asymptotic behavior
\begin{equation}\label{asymp}
\left| \frac{u_m(t)}{\widetilde{u}_m(t)} - 1 \right| = O\left(\frac{\beta cDt}{x_{c_3}}\right), \quad \frac{\beta cDt}{x_{c_3}} \to 0.
\end{equation}
\end{lemma}
\textbf{Proof:}
By applying the integration by parts for the complementary error function, we have
\begin{eqnarray*}
\mbox{erfc}(\xi) &=& \frac{2}{\sqrt{\pi}}\int_\xi^\infty e^{-s^2}\;{\rm d}s \\
&=& \frac{e^{-\xi^2}}{\sqrt{\pi}\xi}\left[ 1- \xi e^{\xi^2}\int_\xi^\infty s^{-2}e^{-s^2}\;{\rm d}s \right]
\end{eqnarray*}
and
\begin{equation*}
\sqrt{\pi}\xi e^{\xi^2}\mbox{erfc}(\xi) \le 1
\end{equation*}
for $\xi>0$.
This together with \eqref{eq:3.3} implies
\begin{eqnarray*}
\left|H_\beta(s) - 1\right|  &\le& \beta\sqrt{\pi cDs} e^{\xi^2} \mathrm{erfc}(\xi) \\
&\le& \frac{2\beta cDs}{x_{c3}+2\beta cDs} \le
\delta t
\end{eqnarray*}
for $0<s<t<T$, where $\delta=\frac{2\beta cD}{x_{c3}}$. 
Then by \eqref{eq:3.2} we obtain
\begin{eqnarray}\label{eq:3.4}
\left|f_{A,\beta}(s) - f_{A,0}(s)\right| &=&  s^{-\frac{3}{2}}e^{-\frac{A}{s}} \left|H_\beta(s) - 1\right| \nonumber\\
&\le&  \delta f_{A,0}(s)
\end{eqnarray}
for $0<s<t<T$. Similarly we have $\left|f_{B,\beta}(s) - f_{B,0}(s)\right|\le \delta f_{B,0}(s)$ for $0<s<t<T$.

Note the definition of $ \widetilde{u}_m(t)$ as in \eqref{tid-um}. By \eqref{eq:3.1}, \eqref{tid-um} and \eqref{eq:3.4}, we obtain the estimation
\begin{eqnarray*}\label{eq:3.6}
&&\left|u_m(t) - \widetilde{u}_m(t)\right| \\
& \le& C(t) \int_0^t \left| f_{A,\beta}(t-s) f_{B,\beta}(s) - f_{A,0}(t-s)f_{B,0}(s) \right| \,{\rm d}s \\
&\le& C(t) \int_0^t \left| f_{A,\beta}(t-s) - f_{A,0}(t-s) \right| f_{B,\beta}(s)  \, {\rm d}s \\
& &+ C(t) \int_0^t  f_{A,0}(t-s)\left| f_{B,\beta}(s) - f_{B,0}(s) \right| \, {\rm d}s \\
&\le & 2\delta t\widetilde{u}_m(t)
\end{eqnarray*}
for $0<t<T$, implying the asymptotic behavior as in \eqref{asymp}. 

The proof is complete.

\begin{lemma}\label{thm-asym2}
Corresponding to the case of $\beta=+\infty$, we set
\begin{equation}\label{tilde-w}
\widetilde{w}_m(t) = \delta^{-2} C(t) \int_0^t \frac{f_{A,0}(t-s)}{t-s} \frac{f_{B,0}(s)}{s}\; {\rm d}s, 
\quad \delta = \frac{2\beta cD}{x_{c3}},
\end{equation}
which is the boundary measurements of \eqref{sun2-4} corresponding to the Dirichlet boundary condition. Then, there holds the following asymptotic behavior
\begin{equation}
\left| u_m(t)-\widetilde{w}_m(t) \right| = O\left(\beta^{-3}\right), \quad \beta \to +\infty.
\end{equation}
\end{lemma}
\textbf{Proof:}
Since
$$
\sqrt{\pi}e^{\xi^2} \mathrm{erfc}(\xi) = \xi^{-1} + O\left(\xi^{-3}\right) \quad \mbox{as} \quad \xi  \to +\infty, 
$$ 
this together with \eqref{eq:3.3} implies
\begin{eqnarray}\label{D1}
&&\left|H_\beta(s) - \delta^{-1} \frac{1}{s}\right| \nonumber\\
&\le& 
\left| \left(1 - \beta \sqrt{ cDs} \xi^{-1}\right) - \frac{x_{c3}}{2\beta cDs}\right| + C \beta \sqrt{ cDs} \xi^{-3} \nonumber \\
&\le& \left|\frac{x_{c3}}{x_{c3}+2\beta cDs}  - \frac{x_{c3}}{2\beta cDs}\right| +
 C\frac{\beta (cDs)^2}{(x_{c3}+2\beta cDs)^3} \\
&\le&
 \frac{x_{c3}^2}{(2\beta cDs)^2} +
 C\frac{\beta (cDs)^2}{(x_{c3}+2\beta cDs)^3} \nonumber\\
&\le& C\beta^{-2} \left(\frac{1 }{s^2}+\frac{1}{s}\right) \nonumber
\end{eqnarray}
for $s>0$, where $C>0$ is independent of $s>0$ and $\beta>0$.
Here constants $C$ may have different values also within the same line.
We remark $\xi \to \infty$ as $\beta \to \infty$. 
Then by \eqref{eq:3.2}
\begin{eqnarray*}
\left|f_{A,\beta}(s) - \delta^{-1} \frac{f_{A,0}(s)}{s}\right| &=&  s^{-\frac{3}{2}}e^{-\frac{A}{s}} \left|H_\beta(s) - \delta^{-1} \frac{1}{s}\right| \nonumber\\
&\le&  C\beta^{-2} \left(\frac{1 }{s^2}+\frac{1}{s}\right) f_{A,0}(s),
\end{eqnarray*}
and
\begin{eqnarray*}
\left|f_{B,\beta}(s) - \delta^{-1} \frac{f_{B,0}(s)}{s}\right| &=&  s^{-\frac{3}{2}}e^{-\frac{B}{s}} \left|H_\beta(s) - \delta^{-1} \frac{1}{s}\right| \nonumber\\
&\le&  C\beta^{-2} \left(\frac{1 }{s^2}+\frac{1}{s}\right) f_{B,0}(s)
\end{eqnarray*}
for $0<s<t<T$.
Then this together with \eqref{eq:3.1}, \eqref{tilde-w} and \eqref{D1} implies that
\begin{eqnarray*}
& &\left|u_m(t) - \widetilde{w}_m(t)\right| \\
&\le& C(t) \int_0^t \left| f_{A,\beta}(t-s) f_{B,\beta}(s) - 
\delta^{-2} \frac{f_{A,0}(t-s)}{t-s}\frac{f_{B,0}(s)}{s} \right| \; {\rm d}s \\
&=& C(t) \int_0^t \left| f_{A,\beta}(t-s) -\delta^{-1} \frac{f_{A,0}(t-s)}{t-s} \right| f_{B,\beta}(s)  \; {\rm d}s \\
&&+ C(t) \int_0^t \delta^{-1} \frac{f_{A,0}(t-s)}{t-s}\left| f_{B,\beta}(s) -\delta^{-1}\frac{f_{B,0}(s)}{s} \right|   \; {\rm d}s \\
&=& O\left(\beta^{-3}\right)
\end{eqnarray*}
as $\beta \to +\infty$. 

Since 
$$
\widetilde{w}_m(t) = O\left(\beta^{-2}\right) \quad \mbox{as} \quad \beta \to +\infty,
$$
we obtain 
$$
u_m(t) \sim \widetilde{w}_m(t) \quad \mbox{as} \quad \beta \to +\infty,
$$
where $\widetilde{w}_m$ is as in \eqref{tilde-w}. The proof is complete.

\section{The peak time of boundary measurements}\label{sec3}

In this section, we will show the expressions of the approximate peak time for TD boundary measurements \eqref{ue-mesure} and \eqref{eq:3.1}, respectively. 

\subsection{The expression of peak time for $u_e$}

We consider the expression of peak time for TD boundary measurements $u_e$ as in \eqref{ue-mesure}. Likewise Lemma \ref{thm-asym}, we first investigate the asymptotic behavior of $u_e$ in the following two limiting cases: 
\begin{equation*}
\beta \sqrt{cDt} \ll 1
\end{equation*} 
and 
\begin{equation*}
\beta \sqrt{cDt} \gg 1.
\end{equation*}
Then, based on the asymptotic expansion of $u_e$, we are able to present the expression of the approximate peak time for $u_e$. By the expression of $u_e$ as in \eqref{ue-mesure}, we have
\begin{equation}
\label{eq:4.1}
u_e(t) = {\widetilde C}(t) t^{-\frac{3}{2}} e^{-{\widetilde A}/{t}} {\widetilde K}_3(t), 
\end{equation}
where
\begin{equation*}
{\widetilde C}(t) :=\frac{ e^{-c\mu_a t}}{4\pi^{3/2} \sqrt{Dc}}, \quad {\widetilde A}:= \frac{\|x-x_s\|^2}{4cD}, 
\end{equation*}
and
\begin{equation*} 
{\widetilde K}_3(t) := {\widetilde K}_3(0,0;t)=2\left(1-\sqrt{\pi}\xi e^{\xi^2}\mbox{erfc}(\xi)\right)
\end{equation*}
with $\xi:=\xi(t) = \beta \sqrt{cDt}$. We have the asymptotic behavior for ${\widetilde K}_3(t)$ that
\begin{equation}\label{eq:4.2}
{\widetilde K}_3(t) = 2 + O \left( \xi \right)
\end{equation}
as $\xi \to 0$, and  
\begin{eqnarray}\label{eq:4.3}
{\widetilde K}_3(t) &=& 2\left(1-\sqrt{\pi}\xi e^{\xi^2}\mbox{erfc}(\xi)\right)  \nonumber\\
 &=&2\xi e^{\xi^2} \int_\xi^\infty s^{-2} e^{-s^2} \, {\rm d}s \nonumber \\
&=& \xi^{-2} - \frac{3}{2}\xi^{-4} + O \left( \xi^{-6} \right)
\end{eqnarray}
as $\xi \to \infty$. Now we are ready to give the expression of approximate peak time of $u_e(t)$ as follows.
\begin{thm}
(1) If $\beta \sqrt{cDT} \ll 1$, the TD boundary measurements $u_e(t)$ uniquely attains the maximum approximately at
\begin{equation}\label{peak1-ue}
t_{\rm peak} =\frac{1}{4\left[c\mu_a - \beta^2cD\right]}
\left( -3 + \sqrt{9 + 16{\widetilde A}\left[c\mu_a - \beta^2cD\right] }\right) .
\end{equation}
(2) If $\beta \sqrt{cDT} \gg 1$, then $u_e(t)$ uniquely attains the maximum approximately at
\begin{equation}\label{peak2-ue}
t_{\rm peak} =\frac{1}{4 c\mu_a}\left( -5 + \sqrt{25 + 16{\widetilde A} c\mu_a  }\right) .
\end{equation}

\end{thm}
{\bf Proof:}
Since 
$$
\frac{d {\widetilde K}_3}{d t} = \left(\frac{1}{2}t^{-1} +  t^{-1}\xi^2 \right){\widetilde K}_3(t)
- t^{-1},
$$
we obtain
\begin{equation*}
\frac{d {\widetilde K}_3}{d t} \sim
\left\{
\begin{array}{ll}
[\beta^2 cD] {\widetilde K}_3(t) &\quad \mbox{if} \quad \xi\ll 1, \vspace{3pt}\\
-t^{-1}{\widetilde K}_3(t) &\quad \mbox{if} \quad \xi\gg 1, \\
\end{array}
\right.
\end{equation*}
by using \eqref{eq:4.2} and \eqref{eq:4.3}. This together with the expression of the derivative of $u_e$
\begin{equation*}
\frac{du_e}{dt}=\left[-c\mu_a -\frac{3}{2}t^{-1} + {\widetilde A}t^{-2}\right]u_e + {\widetilde C}(t) t^{-\frac{3}{2}} e^{-{\widetilde A}/{t}} \frac{d {\widetilde K}_3}{d t}
\end{equation*}
implies that
\begin{equation*}
\frac{du_e}{dt}\sim
\left\{
\begin{array}{ll}
\left[-c\mu_a -\frac{3}{2}t^{-1} + {\widetilde A}t^{-2}+ \beta^2 cD \right]u_e &\;\; \mbox{if} \;\; \xi \ll 1, \vspace{3pt}\\
\left[-c\mu_a -\frac{5}{2}t^{-1} + {\widetilde A}t^{-2} \right]u_e &\;\; \mbox{if} \;\; \xi \gg 1. \\
\end{array}
\right.
\end{equation*}

Finally, computing $\frac{d u_e}{d t}=0$ yields that $u_e(t)$ attains the maximum approximately at $t=t_{\rm peak}$ as in \eqref{peak1-ue} and \eqref{peak2-ue} with respect to the cases $\beta \sqrt{cDt_{\rm peak}} \ll 1$ and $\beta \sqrt{cDt_{\rm peak}} \gg 1$. Further, it is clear that  the peak time is unique.

The proof is complete.

\subsection{The Neumann boundary condition}

We consider the Neumann boundary condition, i.e., $\beta=0$ in \eqref{excitation-Ue} and \eqref{sun2-4}. Then we investigate the peak time for TD boundary measurements (i.e., $\widetilde u_m$ as in \eqref{tid-um}). We first consider the existence of peak time for $\widetilde u_m$, which is to prove that there exists one time point such that $\frac{d{\widetilde u_m}}{dt}(t)=0$. The precise statement is given by the following theorem.

\begin{thm}\label{existence}
\label{thm:3.2}
There exist $0<t_1<t_2$ such that 
\begin{equation}\label{eq:3.7-1}
\frac{d \widetilde{u}_m }{d t} >0 \quad \mbox{for} \quad  t <t_1,
\end{equation}
and
\begin{equation}\label{eq:3.7}
\frac{d \widetilde{u}_m }{d t} <0 \quad \mbox{for} \quad  t>t_2.
\end{equation}
\end{thm}
{\bf Proof:} Observe that 
\begin{eqnarray}\label{symmetry}
\widetilde{u}_m(t)&=&C(t) \left[\int_0^{t/2}f_A(t-s)f_B(s)\,{\rm ds}\right.\\
& &+\left.\int_0^{t/2}f_B(t-s)f_A(s)\,{\rm ds}\right].
\end{eqnarray}
Then we have
\begin{eqnarray}\label{eq:3.8}
 \frac{d \widetilde{u}_m}{d t} &=& C(t)
 \left[\int_0^\frac{t}{2} \Big(\eta_A(t-s)-c\mu_a\Big) f_A(t-s) f_B(s) \; {\rm d}s\right. \nonumber \\ 
& & + \left.\int_0^\frac{t}{2} \Big(\eta_B(t-s)-c\mu_a\Big) f_B(t-s) f_A(s) \; {\rm d}s \right] \nonumber \\ 
&& +C(t) f_A\left(\frac{t}{2}\right) f_B\left({\frac{t}{2}}\right)
\end{eqnarray}
for $t>0$, where $f_A := f_{A,0}$, $f_B:=f_{B,0}$ are as in \eqref{eq:3.2} and
\begin{equation*}
\eta_A(s) := -\frac{3}{2}s^{-1}+As^{-2}, \;\;
\eta_B(s) := -\frac{3}{2}s^{-1}+Bs^{-2}
\end{equation*}
for $t/2\le s\le t$.
Then we have for $t<t_A$ that
\begin{equation*}
\eta_A(t-s) = -\frac{3}{2}(t-s)^{-1}+A(t-s)^{-2} > c\mu_a 
\end{equation*}
for $0<s\le\frac{t}{2}$, where $t_A$ is the positive solution of
\begin{equation*}
-\frac{3}{2}t^{-1}+At^{-2} = c\mu_a.
\end{equation*}
That is
\begin{equation*}
t_A := \frac{-3+\sqrt{9+16c\mu_a A}}{4c\mu_a}.
\end{equation*}

Let $t_B$ is the positive solution of
\begin{equation*}
-\frac{3}{2}t^{-1}+Bt^{-2} = c\mu_a.
\end{equation*}
We take $t_1:=\min\{t_A,t_B\}>0$.
Then by \eqref{eq:3.8} we have 
\begin{equation}\label{eq:3.9}
\frac{d \widetilde{u}_m} {dt} > 0 \quad \mbox{for} \quad  0<t< t_1,
\end{equation}
implying \eqref{eq:3.7-1} is established.

On the other hand, for $t/2>2A/3$, we have
\begin{equation*}
\eta_A(t-s) < 0, \quad f_A(t-s)>f_A(t) 
\end{equation*}
for $0<s<t/2$.
This implies
\begin{eqnarray}
\label{eq:3.10}
&& \int_0^\frac{t}{2} \Big(\eta_A(t-s)-c\mu_a\Big) f_A(t-s) f_B(s) \; {\rm d}s +
\frac{1}{2}f_A\Big(\frac{t}{2}\Big) f_B\Big(\frac{t}{2}\Big) \nonumber\\
&< &-c\mu_a\int_0^\frac{t}{2} f_A(t-s)f_B(s) \; {\rm d}s +
\frac{1}{2}f_A\Big(\frac{t}{2}\Big)f_B\Big(\frac{t}{2}\Big) \nonumber\\
&<& f_A(t)\left[ -c\mu_a \int_0^\frac{t}{2}f_B(s) \; {\rm d}s + \sqrt{2}f_B\Big(\frac{t}{2}\Big) \right]  
\end{eqnarray}
for $t>4A/3$. 
Since
\begin{eqnarray*}
\int_0^\frac{t}{2} f_B(s) \; {\rm d}s &=& \int_0^\infty f_B(s) \; {\rm d}s - \int_\frac{t}{2}^\infty f_B(s) \; {\rm d}s \\
&\ge& m_B -  \int_\frac{t}{2}^\infty s^{-\frac{3}{2}} \; {\rm d}s = m_B-2\sqrt{2}t^{-\frac{1}{2}}
\end{eqnarray*}
with $m_B=\int_0^\infty f_B(s)\,{\rm ds}=\sqrt{\pi/B}$ for $t>0$, we obtain
\begin{eqnarray}\label{eq:3.11}
&&-c\mu_a \int_0^\frac{t}{2} f_B(s) \; {\rm d}s + \sqrt{2} f_B\Big(\frac{t}{2}\Big) \nonumber\\
&<&-c\mu_a \left(m_B-2\sqrt{2}t^{-\frac{1}{2}}\right) + 4 t^{-\frac{3}{2}} \nonumber\\
&<&-c\mu_a m_B + t^{-\frac{1}{2}} \left( 2\sqrt{2}c\mu_a + \frac{3}{B} \right) <0
\end{eqnarray}
for $t>\max \left\{ \frac{4B}{3}, \left(\frac{2\sqrt{2}c\mu_a + \frac{3}{B}}{c\mu_a m_B}\right)^2 \right\}$.

Set 
\begin{equation*}
t_2 := \max \left\{ \frac{4A}{3}, \; \frac{4B}{3}, \;
\left(\frac{2\sqrt{2}c\mu_a + \frac{3}{A}}{c\mu_a m_A}\right)^2, \;
\left(\frac{2\sqrt{2}c\mu_a + \frac{3}{B}}{c\mu_a m_B}\right)^2 \right\}
\end{equation*}
with $m_A =\int_0^\infty f_A(s)\,{\rm ds}= \sqrt{\frac{\pi}{A}}$. Then by \eqref{eq:3.8}, \eqref{eq:3.10} and \eqref{eq:3.11} we have
\begin{equation}
\frac{d \widetilde{u}_m} {dt} < 0 \quad \mbox{for} \quad  t> t_2,
\end{equation}
implying \eqref{eq:3.7} is established. The proof is complete.
\vspace{3pt}

In Theorem \ref{existence}, we only proved the existence of peak time of $\widetilde{u}_m$ for general $A>0$ and $B>0$, without giving its uniqueness. However, for the case of $A=B$, we are able to show the uniqueness of peak time and present its explicit formula.

\begin{thm}\label{uniqueness}
Suppose $A=B$. 
Then there holds 
\begin{equation}\label{um-first}
\widetilde{u}_m(t) = 2 C(t) \sqrt{\frac{\pi}{A}} \exp\left(-\frac{4A}{t} \right) t^{-\frac{3}{2}},
\end{equation}
where $C(t)$ and $A$ are as in \eqref{parameter}. Further, $\widetilde{u}_m$ admits the unique peak time 
\begin{equation}\label{peak_exa}
t_{\rm peak} = \frac{1}{4c\mu_a} \left( -3+\sqrt{9+64c\mu_a A}\right).
\end{equation}
\end{thm}
{\bf Proof:}
By \eqref{symmetry}, we have
\begin{eqnarray*}\label{um-AB}
\widetilde{u}_m(t)&=& C(t) 
\int_0^t \big[(t-s)s\big]^{-\frac{3}{2}}  e^{-\frac{A}{t-s}} e^{-\frac{B}{s}} \,{\rm d}s \nonumber \\
&=& C(t) (I_1+I_2),
\end{eqnarray*}
where
\begin{eqnarray}
I_1=I_1(t)&:=& \int_0^\frac{t}{2} \big[(t-s)s\big]^{-\frac{3}{2}}e^{-\frac{Bt}{(t-s)s}} e^{-\frac{A-B}{t-s}} \, {\rm d}s,\label{I1}\\
I_2=I_2(t)&:=& \int_0^\frac{t}{2}  \big[(t-s)s\big]^{-\frac{3}{2}}e^{-\frac{At}{(t-s)s}} e^{-\frac{B-A}{t-s}}\,{\rm d}s.\label{I2}
\end{eqnarray}
Since $A=B$, it is easy to see that
\begin{equation*}
I_1=I_2=\int_0^\frac{t}{2} \big[(t-s)s\big]^{-\frac{3}{2}}e^{-\frac{At}{(t-s)s}} \, {\rm d}s.
\end{equation*}
By the change of variable $\sigma=s/t$, there holds
\begin{equation*}\label{It2}
I_1 = t^{-2} \int_0^{1/2} [\sigma(1-\sigma)]^{-3/2} e^{-\frac{At^{-1}}{\sigma(1-\sigma)}} \, {\rm d}\sigma.
\end{equation*}
We transform the integration variable $\sigma$ to $z=(\sigma(1-\sigma))^{-1}$ which transforms $0\leq \sigma \leq 1/2$ to $4\leq z<\infty$, and $\sigma$ is given as $\sigma=\big(1-\sqrt{1-4z^{-1}}\big)/2$. Then, further transforming $z$ to $\zeta=z-4$, which yields the following
\begin{equation*}
I_1 = \exp\left(-\frac{4A}{t} \right) t^{-2} \int_0^\infty \zeta^{-1/2} e^{-At^{-1}\zeta}\, {\rm d}\zeta.
\end{equation*}
By straight computation, we have 
\begin{equation*}
I_1=I_2 = \sqrt{\frac{\pi}{A}} \exp\left(-\frac{4A}{t} \right)  t^{-\frac{3}{2}}.
\end{equation*}
This together with \eqref{um-AB} implies 
\begin{equation*}
\widetilde{u}_m(t) = 2C(t) \sqrt{\frac{\pi}{A}} \exp\left(-\frac{4A}{t} \right)  t^{-\frac{3}{2}},
\end{equation*}
and
\begin{equation*}
\frac{d \widetilde{u}_m}{dt}  =  \widetilde{u}_m(t) \left[-c\mu_a -\frac{3}{2}t^{-1} +4At^{-2}\right].
\end{equation*}
Then, by computing $\frac{d \widetilde{u}_m}{dt}=0$ we can see that $\widetilde{u}_m(t)$ attains the maximum at the time $t_{\rm peak}$ as in \eqref{peak_exa}.

The proof is complete.
\vspace{3pt}

We next consider the peak time of $\widetilde{u}_m(t)$ for the case $A\neq B$ but $|A-B|$ is small enough. 
Set 
\begin{equation}
\label{eq:3.16}
\widetilde{v}_m(t) := C(t) \left(\sqrt{\frac{\pi}{A}} + \sqrt{\frac{\pi}{B}} \right)
\exp\left(-\frac{2A+2B}{t} \right)  t^{-\frac{3}{2}}.
\end{equation}
We will prove ${\widetilde u}_m\sim\widetilde{v}_m$ as ${|A-B|}/{t} \to 0$ and give the expression of approximate peak time for ${\widetilde u}_m$ with respect to the case $|A-B|$ is very small, which is stated as the following theorem.
\begin{thm}\label{thmAnB}
Let ${\widetilde u}_m$ and ${\widetilde v}_m$ be as in \eqref{tid-um} and \eqref{eq:3.16}, respectively.
Then 
\begin{equation}
\label{eq:3.17}
\left|\frac{{\widetilde u}_m(t)}{{\widetilde v}_m(t)} -1\right| = O\left(\frac{|A-B|}{t}\right), \quad  \frac{|A-B|}{t} \to 0.
\end{equation}
Further, if $|A-B|/t_{\rm peak}\ll 1$, the peak time of ${\widetilde u}_m(t)$ can be approximated by
\begin{equation}\label{peak_app}
t_{\rm peak} \approx \frac{1}{4c\mu_a} \left( -3+\sqrt{9+32c\mu_a (A+B)}\right).
\end{equation}
\end{thm}
{\bf Proof:}
By denoting $\gamma:=(A-B)t^{-1}$ and  a transformation of integration variable for \eqref{I1} and \eqref{I2}, we have
\begin{eqnarray}
I_1 &=& t^{-2}\int_0^{\frac{1}{2}} (\sigma(1-\sigma))^{-\frac{3}{2}}
e^{-\frac{Bt^{-1}}{\sigma(1-\sigma)}} e^{-\frac{\gamma}{1-\sigma}}\, {\rm d}\sigma, \label{II1}\\
I_2 &=& t^{-2}\int_0^{\frac{1}{2}} (\sigma(1-\sigma))^{-\frac{3}{2}}
e^{-\frac{At^{-1}}{\sigma(1-\sigma)}} e^{-\frac{-\gamma}{1-\sigma}}\, {\rm d}\sigma.  \label{II2}
\end{eqnarray}
Note that 
\begin{eqnarray*}
&&t^{-2}\int_0^{\frac{1}{2}} (\sigma(1-\sigma))^{-\frac{3}{2}}
e^{-\frac{Bt^{-1}}{\sigma(1-\sigma)}} e^{-2\gamma} \, {\rm d}\sigma \\
&=&\sqrt{\frac{\pi}{B}}
\exp\left(-\frac{2A+2B}{t} \right)  t^{-\frac{3}{2}},
\end{eqnarray*}
and
\begin{eqnarray*}
&&t^{-2}\int_0^{\frac{1}{2}} (\sigma(1-\sigma))^{-\frac{3}{2}}
e^{-\frac{At^{-1}}{\sigma(1-\sigma)}} e^{2\gamma} \, {\rm d}\sigma \\
&=&\sqrt{\frac{\pi}{A}}
\exp\left(-\frac{2A+2B}{t} \right)  t^{-\frac{3}{2}}.
\end{eqnarray*}

Now we are ready to show the asymptotic behavior as in \eqref{eq:3.17}. By the mean value theorem for $e^{-\frac{\gamma}{1-\sigma}}$ between $\sigma$ and $\frac{1}{2}$, there exists $\sigma < \theta < \frac{1}{2}$ such that
\begin{eqnarray*}
\left|  e^{-\frac{\gamma}{1-\sigma}} - e^{-2\gamma} \right| 
&=& e^{-\frac{\gamma}{1-\theta}} \frac{|\gamma|}{(1-\theta)^2}\left|\sigma-\frac{1}{2}\right| \\ 
&\le& 4 |\gamma| \left|\sigma-\frac{1}{2}\right| \times \max\{e^{-\gamma}, e^{-2\gamma}\}
\end{eqnarray*}
for $ 0<\sigma<\frac{1}{2}$.
This together with \eqref{II1} implies
\begin{eqnarray}\label{eq:3.20}
&& \left| I_1(t) - \sqrt{\frac{\pi}{B}}\exp\left(-\frac{2A+2B}{t} \right)  t^{-\frac{3}{2}} \right| \nonumber \\
&=& \left| I_1(t) - t^{-2}\int_0^{\frac{1}{2}} (\sigma(1-\sigma))^{-\frac{3}{2}}
e^{-\frac{Bt^{-1}}{\sigma(1-\sigma)}} e^{-2\gamma} \, {\rm d}\sigma \right| \nonumber \\
&\le& 4 |\gamma| \max\{e^{-\gamma}, e^{-2\gamma}\}   \nonumber \\
&&\times \left(t^{-2} \int_0^{\frac{1}{2}} (\sigma(1-\sigma))^{-\frac{3}{2}}
e^{-\frac{Bt^{-1}}{\sigma(1-\sigma)}}\left|\sigma-\frac{1}{2}\right|  \, {\rm d}\sigma\right)  \nonumber \\
&\le& 2  |\gamma| \max\{e^{-\gamma}, e^{-2\gamma}\} 
\left( t^{-2}\int_0^{\frac{1}{2}} (\sigma(1-\sigma))^{-\frac{3}{2}}
e^{-\frac{Bt^{-1}}{\sigma(1-\sigma)}} \, {\rm d}\sigma \right) \nonumber \\
&=& 2  |\gamma| \max\{1, e^\gamma \} \sqrt{\frac{\pi}{B}} 
\exp\left(-\frac{2A+2B}{t} \right)  t^{-\frac{3}{2}}.
\end{eqnarray}
Similarly, we have the estimation for $I_2(t)$ that
\begin{eqnarray}\label{eq:3.20-2}
&& \left| I_2(t) - \sqrt{\frac{\pi}{A}}\exp\left(-\frac{2A+2B}{t} \right)  t^{-\frac{3}{2}} \right| \nonumber \\
&\le&  2  |\gamma| \max\{1, e^{\gamma} \} \sqrt{\frac{\pi}{A}} 
\exp\left(-\frac{2A+2B}{t} \right)  t^{-\frac{3}{2}}.
\end{eqnarray}
Then, by \eqref{eq:3.20} and \eqref{eq:3.20-2} we immediately have the estimation 
\begin{eqnarray*}
&& \left| (I_1+I_2) - \left(\sqrt{\frac{\pi}{A}}+\sqrt{\frac{\pi}{B}}\right)\exp\left(-\frac{2A+2B}{t} \right)  t^{-\frac{3}{2}} \right| \nonumber \\
&\le& 2  |\gamma| e^{|\gamma|}  \left(\sqrt{\frac{\pi}{A}}+\sqrt{\frac{\pi}{B}}\right) 
\exp\left(-\frac{2A+2B}{t} \right)  t^{-\frac{3}{2}},
\end{eqnarray*}
implying that
\begin{equation*}
| \widetilde u_m(t) - \widetilde v_m(t)| = \widetilde v_m(t) O\left(|\gamma| e^{|\gamma|} \right)
\end{equation*}
as $\gamma \to 0$.
By the expression of $\widetilde{v}_m$ as in \eqref{eq:3.16}, we obtain 
\begin{equation*}
\frac{d \widetilde{v}_m}{dt} =  \widetilde{v}_m(t) \left[-c\mu_a -\frac{3}{2}t^{-1} +2(A+B)t^{-2}\right].
\end{equation*}
Finally, by computing $\frac{d \widetilde{v}_m}{dt} =0$ we can see that $\widetilde v_m$ attains the maximum at the time as in \eqref{peak_app}. 

The proof is complete.
\vspace{3pt}

\begin{remark}
Based on the asymptotic behavior of $u_m$ in above Lemma \ref{thm-asym}, the peak time of $u_m$ as in \eqref{eq:3.1} can be approximated by \eqref{peak_app} if $\beta cDt_{\rm peak}/x_{c_3} \ll 1$ and $|A-B|/t_{\rm peak}\ll 1$.
\end{remark}

\subsection{The Dirichlet boundary condition}

Here we consider the Dirichlet boundary condition, i.e., $\beta=+\infty$ in \eqref{excitation-Ue} and \eqref{sun2-4}. Then we study the peak time of TD boundary measurements  (i.e., $\widetilde{w}_m$ as in \eqref{tilde-w}). By the similar arguments as in Theorem~\ref{uniqueness} and Theorem~\ref{thmAnB}, we will show the expression of peak time for $\widetilde {w}_m(t)$.

\begin{thm}\label{beta-infty}
Suppose $A=B$. 
Then there holds 
\begin{equation}\label{exp_w}
\widetilde{w}_m(t) = 2 \delta^{-2} C(t) \sqrt{\frac{\pi}{A}} \exp\left(-\frac{4A}{t} \right) \left(4 t^{-\frac{7}{2}} 
+ \frac{1}{2A}t^{-\frac{5}{2}} \right),
\end{equation}
where $C(t)$ and $A$ are as in \eqref{parameter}. Further, $\widetilde{w}_m$ admits unique peak time
\begin{equation}\label{peak_equation}
t_{\rm peak}=\frac{1}{6c\mu_a}\left\{-5-16c\mu_aA+\frac{M(A)}{[N(A)]^{1/3}}-[N(A)]^{1/3}\right\},
\end{equation}
where
$$
M(A):=-25+128\rho_A-256\rho_A^2
$$
and
\begin{eqnarray*}
N(A)&:=& 125-960\rho_A-6528\rho_A^2+4096\rho_A^3 \\
&&+48\sqrt{6}\sqrt{-175\rho_A^2+1488\rho_A^3+1248\rho_A^4-2048\rho_A^5}
\end{eqnarray*}
with $\rho_A:=c\mu_aA$.
\end{thm}
{\bf Proof:}
By \eqref{tilde-w}, we have
\begin{eqnarray}\label{wm-AB}
\widetilde{w}_m(t)&=& \delta^{-2} C(t) 
\int_0^t \big[(t-s)s\big]^{-\frac{5}{2}}  e^{-\frac{A}{t-s}} e^{-\frac{B}{s}} \,{\rm d}s \nonumber \\
&=& \delta^{-2} C(t) ({\widetilde I}_1+{\widetilde I}_2),\label{tilde-I1I2}
\end{eqnarray}
where
\begin{eqnarray}
{\widetilde I}_1={\widetilde I}_1(t)&:=& \int_0^\frac{t}{2} \big[(t-s)s\big]^{-\frac{5}{2}}e^{-\frac{Bt}{(t-s)s}} e^{-\frac{A-B}{t-s}} \, {\rm d}s,\\
{\widetilde I}_2={\widetilde I}_2(t)&:=& \int_0^\frac{t}{2}  \big[(t-s)s\big]^{-\frac{5}{2}}e^{-\frac{At}{(t-s)s}} e^{-\frac{B-A}{t-s}}\,{\rm d}s.
\end{eqnarray}

Likewise Theorem \ref{uniqueness}, we have by $A=B$ that
\begin{eqnarray*}
{\widetilde I}_1={\widetilde I}_2&= &\int_0^\frac{t}{2} \big[(t-s)s\big]^{-\frac{5}{2}}e^{-\frac{At}{(t-s)s}} \, {\rm d}s \\
&=& t^{-4} \int_0^{1/2} [\sigma(1-\sigma)]^{-5/2} e^{-\frac{At^{-1}}{\sigma(1-\sigma)}} \, {\rm d}\sigma \\
&=&\exp\left(-\frac{4A}{t} \right) t^{-4} \int_0^\infty (4+\zeta) \zeta^{-1/2} e^{-At^{-1}\zeta}\, {\rm d}\zeta.
\end{eqnarray*}
This together with \eqref{tilde-I1I2} implies \eqref{exp_w} and
\begin{equation*}
\begin{split}
\frac{d \widetilde{w}_m}{dt} =
&2 \delta^{-2} C(t) \sqrt{\frac{\pi}{A}} \exp\left(-\frac{4A}{t} \right) \\
&\times \left[-\frac{c\mu_a}{2A}t^3-\left(4c\mu_a+\frac{5}{4A}\right)t^2-12t+16A\right].
\end{split}
\end{equation*}
Then, by computing $\frac{d \widetilde{w}_m}{d t}=0$ we can see that $\widetilde{w}_m(t)$ attains the maximum at the time $t_{\rm peak}$ as in \eqref{peak_equation}, i.e., \eqref{peak_equation} is the unique positive zero of the following cubic polynomial given as
\begin{equation}
p(t):=-\frac{c\mu_a}{2A}t^3-\left(4c\mu_a+\frac{5}{4A}\right)t^2-12t+16A
\end{equation}
because $p'(t) <0$ for $t>0$.
Then the proof is complete.
\vspace{3pt}

\begin{remark}\label{rem-w}
If $t/A\ll 1$, by \eqref{exp_w} we have
\begin{equation}
\widetilde{w}_m(t) \sim 8 \delta^{-2} C(t) \sqrt{\frac{\pi}{A}} \exp\left(-\frac{4A}{t} \right) t^{-\frac{7}{2}}.
\end{equation}
This yields that $\widetilde{w}_m$ admits unique approximate peak time
\begin{equation}\label{peak2_beta_infty}
t_{\rm peak} \sim \frac{1}{4c\mu_a} \left( -7+\sqrt{49+64c\mu_a A}\right)
\end{equation}
under the condition $t_{\rm peak}/A\ll 1$.
\end{remark}
\vspace{3pt}

We next consider the peak time of $\widetilde{w}_m(t)$ for the case $A\neq B$ but $|A-B|$ is small enough. 
Set 
\begin{eqnarray*}
{\widetilde z}_m(t) &:=&\delta^{-2} C(t) \exp\left(-\frac{2A+2B}{t}\right)
\\ 
&&\times \left(4\left[\sqrt{\frac{\pi}{A}}
+\sqrt{\frac{\pi}{B}}\right]t^{-\frac{7}{2}} + \left[\frac{1}{2A}\sqrt{\frac{\pi}{A}}+\frac{1}{2B}\sqrt{\frac{\pi}{B}} \right]t^{-\frac{5}{2}} \right).
\end{eqnarray*}
Likewise Theorem~\ref{thmAnB}, we can have the asymptotic behavior of ${\widetilde w}_m$. Then we arrive at the expression of peak time for ${\widetilde w}_m$ under the condition $|A-B|$ is small enough. Since the same argument in Theorem~\ref{thmAnB} still works in our setting, we will state the following conclusion and omit the proof here.
\begin{thm}\label{zm-wm}
There holds
\begin{equation}\label{w_app}
\left|\frac{{\widetilde w}_m(t)}{{\widetilde z}_m(t)} -1\right| = O\left(\frac{|A-B|}{t}\right), \quad  \frac{|A-B|}{t} \to 0.
\end{equation}
Then, if $A\neq B$ but $|A-B|/t \ll 1$, the expression of peak time for ${\widetilde w}_m(t)$ can be approximately given as in \eqref{peak_equation} by replacing $A$ by $(A+B)/2$.
\end{thm}

We will give a remark before closing this subsection.
\begin{remark}
Together Remark \ref{rem-w}, Theorem \ref{zm-wm} with the asymptotic behavior of $u_m$ in Lemma \ref{thm-asym2}, we can approximate the peak time of $u_m$ as in \eqref{eq:3.1} by 
\begin{equation}
t_{\rm peak} \sim \frac{1}{4c\mu_a} \left( -7+\sqrt{49+32c\mu_a (A+B)}\right)
\end{equation}
if $\beta \to \infty$ provided that $t_{\rm peak}/A \ll 1$ and $|A-B|/t_{\rm peak}\ll 1$.
\end{remark}

\subsection{The Robin boundary condition}
\label{robin}

Now we consider the Robin boundary condition, i.e., $\beta$ is a positive constant in \eqref{excitation-Ue} and \eqref{sun2-4}. Then we consider the peak time for $u_m(t)$ as in \eqref{eq:3.1}. We do that basically based on the asymptotic expansion of the complementary error function given as 
\begin{equation}\label{asy-err}
\sqrt{\pi}\xi{e^{\xi^2}} \mathop{\mathrm{erfc}}(\xi)\sim 1,\quad \xi\to\infty.
\end{equation} 
\begin{lemma}\label{lemRobin}
Suppose
\begin{equation}\label{robin:01}
\xi=\xi(t)=\frac{x_{c3}+2\beta cDt}{\sqrt{4cDt}}\gg1,\quad 0<t<T,
\end{equation}
and
\begin{equation}\label{robin:02}
A+B\gg|A-B|. 
\end{equation}
Then, there holds
\begin{eqnarray*}
u_m&\sim&
\frac{2C(t)}{(1+\delta t)t^2}e^{-\frac{2(A+B)}{t}}
\\
&&\times
\left[\sqrt{\pi}\alpha^{-1/2}+\pi(4-\gamma)\gamma^{-1/2}e^{\alpha\gamma}\mathop{\mathrm{erfc}}(\sqrt{\alpha\gamma})\right],
\end{eqnarray*}
where $\delta = \frac{2\beta cD}{x_{c3}}$ and $\gamma$ is given as \eqref{GAM}.
\end{lemma}
{\bf Proof:}
With the assumption \eqref{robin:01}, we have the asymptotic expansion \eqref{asy-err} for any $0<t\leq T$. With the assumption \eqref{robin:02}, there holds
\begin{eqnarray*}
e^{-\frac{A}{t-s}}e^{-\frac{B}{s}}
&=&
\exp\left[-\frac{t}{2(t-s)s}\left(A+B-(A-B)\left(1-\frac{2s}{t}\right)\right)\right]
\nonumber \\
&\sim&
e^{-\frac{(A+B)t}{2(t-s)s}},\quad 0<s<t.
\end{eqnarray*}
Then, by \eqref{eq:3.1}, we can express $u_m(t)$ as
\begin{equation}\label{asy-mas}
u_m(t)\sim u_{m,{\rm as}}(t).
\end{equation}
Here,
\begin{equation*}
u_{m,{\rm as}}(t)=\frac{C(t)}{2+\delta t}\int_0^t\frac{e^{-\frac{(A+B)t}{2(t-s)s}}}{[(t-s)s]^{3/2}}\left(\frac{1}{1+\delta (t-s)}+\frac{1}{1+\delta s}\right)\,{\rm d}s.
\end{equation*}
By changing variables as $\sigma=s/t$, $z=1/[\sigma(1-\sigma)]$, and then $\zeta=z-4$, we arrive at
\begin{equation*}
u_{m,{\rm as}}(t)=
\frac{2C(t)}{(1+\delta t)t^2}e^{-\frac{2(A+B)}{t}}\int_0^{\infty}e^{-\alpha\zeta}\zeta^{-1/2}\frac{\zeta+4}{\zeta+\gamma}\,{\rm d}\zeta,
\end{equation*}
where
\begin{equation}\label{GAM}
\alpha=\alpha(t)=\frac{A+B}{2t}>0,\quad\gamma=\gamma(t)=4+\frac{(\delta t)^2}{1+\delta t}>0.
\end{equation}
By straight computations, we have
\begin{eqnarray*}
&&
\int_0^{\infty}e^{-\alpha\zeta}\zeta^{-1/2}\frac{\zeta+4}{\zeta+\gamma}\,{\rm d}\zeta\\
&=&
\int_0^{\infty}e^{-\alpha\zeta}\zeta^{-1/2}\,{\rm d}\zeta+(4-\gamma)
\int_0^{\infty}e^{-\alpha\zeta}\frac{\zeta^{-1/2}}{\zeta+\gamma}\,{\rm d}\zeta
\\
&=&
\sqrt{\pi}\alpha^{-1/2}+(4-\gamma)\gamma^{-1/2}e^{\alpha\gamma}J(\alpha\gamma),
\end{eqnarray*}
where
\[
J(a):=\int_{-\infty}^{\infty}\frac{e^{-a(s^2+1)}}{s^2+1}\,{\rm d}s,\quad a>0.
\]
We note that $J(\infty)=0$ and $J'(a)=-e^{-a}\sqrt{\pi/a}$. Thus,
\begin{eqnarray*}
\int_a^{\infty}J'(u)\,{\rm d}u
&=&
-J(a)
\\
&=&
-\sqrt{\pi}\int_a^{\infty}e^{-u}u^{-1/2}\,{\rm d}u=-\pi\mathop{\mathrm{erfc}}(\sqrt{a}).
\end{eqnarray*}
Therefore,
\begin{eqnarray*}
u_{m,{\rm as}}(t)
&=&
\frac{2C(t)}{(1+\delta t)t^2}e^{-\frac{2(A+B)}{t}}
\\
&&\times
\left[\sqrt{\pi}\alpha^{-1/2}+\pi(4-\gamma)\gamma^{-1/2}e^{\alpha\gamma}\mathop{\mathrm{erfc}}(\sqrt{\alpha\gamma})\right].
\end{eqnarray*}

The proof is complete.
\vspace{3pt}

Now we give the expression of approximate peak time for $u_m(t)$, which is stated as the following theorem. 

\begin{thm}\label{peak-robin}
Suppose $\frac{2\beta cDt_{\rm peak}}{x_{c3}}\gg1$, $\frac{A+B}{2t_{\rm peak}}\gg1$ and the conditions in Lemma \ref{lemRobin} are established. Then, $u_m(t)$ admits the approximate peak time 
\begin{equation*}
t_{\rm peak} \sim \frac{1}{4c\mu_a} \left( -7+\sqrt{49+32c\mu_a (A+B)}\right).
\end{equation*}
\end{thm}
{\bf Proof:}
By the assumptions
\begin{equation*}
\frac{2\beta cDt_{\rm peak}}{x_{c3}} = \delta t_{\rm peak}\gg1,\quad
\alpha(t_{\rm peak})=\frac{A+B}{2t_{\rm peak}}\gg1,
\end{equation*}
we have
\begin{equation}
\gamma(t_{\rm peak})\sim \delta t_{\rm peak},\quad
\sqrt{\alpha(t_{\rm peak})\gamma(t_{\rm peak})}\gg1.
\end{equation}
Since $\sqrt{\alpha\gamma}$ is large, there holds $e^{\alpha\gamma}\mathop{\mathrm{erfc}}(\sqrt{\alpha\gamma})\sim1/\sqrt{\pi\alpha\gamma}$. Then, at $t\approx t_{\rm peak}$, by Lemma \ref{lemRobin} we have
\begin{eqnarray*}
u_m(t)
&\sim&
\frac{8\sqrt{\pi}C(t)}{\delta t^3}e^{-\frac{2(A+B)}{t}}\alpha^{-1/2}\gamma^{-1}
\\
&\sim&
\frac{x_{c3}^2}{4\sqrt{2}\pi^{5/2}\beta^2D^4c^3\sqrt{A+B}}
t^{-7/2}e^{-c\mu_at}e^{-\frac{2(A+B)}{t}}.
\end{eqnarray*}
Hence,
\begin{eqnarray*}
\frac{d}{dt}u_m(t)
&\sim&
\frac{-x_{c3}^2}{4\sqrt{2}\pi^{5/2}\beta^2D^4c^3\sqrt{A+B}}
t^{-\frac{11}{2}}e^{-c\mu_at}e^{-\frac{2(A+B)}{t}}
\\
&&\times
\left(c\mu_at^2+\frac{7}{2}t-2(A+B)\right).
\end{eqnarray*}
By computing $\frac{d}{dt}u_m(t)=0$, we obtain
\begin{equation}\label{am-peak}
t_{\rm peak}
\sim
\frac{1}{4c\mu_a}\left(-7+\sqrt{49+32c\mu_a(A+B)}\right),
\end{equation}
which is the unique positive solution of following equation 
$$
c\mu_at^2+\frac{7}{2}t-2(A+B)=0.
$$

The proof is complete.
\vspace{3pt}

In Theorem \ref{peak-robin}, under some assumptions for physical parameters, we showed the expression of approximate time for TD boundary measurements with Robin boundary condition. Here we give a remark to show the limiting assumptions in above theorem are not strict in practice.

\begin{remark}\label{physical-par}
The assumption \eqref{robin:01} can be achieved for typical physiological parameters. For example, we set 
\begin{equation}\label{exa-para}
D\sim\frac{1}{3}\,{\rm mm},\quad c\sim 200\,{\rm mm}/{\rm ns}, \quad \beta\sim0.5\,{\rm mm}^{-1},
\end{equation}
which are typical values of biological tissues. Hereafter the unit of length is mm. In this case, there holds
\begin{equation*}
\xi=\xi(t)=\frac{x_{c3}+2\beta cDt}{\sqrt{4cDt}}\gg1,\quad 0<t< T
\end{equation*}
for $x_{c3}\sim 10$. Further, if we set $x=(-20,-20,0)$, $x_s=(20,20,0)$ and $x_c=(0,0,10)$, the assumption $\alpha(t_{\rm peak})\gg1$ can be satisfied. In this case, we have
\begin{eqnarray*}
&&
\alpha(t_{\rm peak})=
\frac{A+B}{2t_{\rm peak}}=\frac{\|x-x_c\|^2+\|x_s-x_c\|^2}{8cDt_{\rm peak}}=
\frac{225}{cDt_{\rm peak}},
\\
&&
\sqrt{\alpha(t_{\rm peak})\gamma(t_{\rm peak})}\sim
\sqrt{\beta\frac{\|x-x_c\|^2+\|x_s-x_c\|^2}{4x_{c3}}}\approx4.7.
\end{eqnarray*}
\end{remark}
\section{Conclusion and Remark}\label{sec4}

In this paper, we considered the TD boundary measurements, which means that the measured data for each given source and detector is a TPSF. One can expect that the TPSF corresponding to a single point target always admits a unique maximum from the physics intuition.  We have mathematically investigated this physical phenomenon in this paper. By analyzing the asymptotic behavior of the TPSF, we proved the unique existence of the peak time for TD boundary measurements.

The peak time is an important property of TPSF. The local data types around the peak time are more robust to noise than other data types, and should provide enhanced information to the related inverse problems such as DOT and FDOT\cite{J.Riley07}. Further, the recovery of unknown information from the measured data around peak time is a robust and efficient approach\cite{D. Hall04}. However, the explicit relations between the peak time and the unknowns have not been discussed yet. In this paper, we explicitly give  an approximate expression of the peak time for both the boundary excitation measurements $u_e$ as in \eqref{ue-mesure} and the boundary emission measurements $u_m$ as in \eqref{mesure}. The expression of approximate peak time for $u_e$ clearly shows the relationship between the peak time and the important information of medium such as the absorption coefficient $\mu_a$ and the diffusion coefficient $D$. The expression of approximate peak time for $u_m$ shows not only the dependence of peak time on $(\mu_a, D)$ but also the connection between the peak time and the position of an unknown point target. The above identities connecting the peak time of measurements and the unknowns in related inverse problems provide the possibility of constructing an effective and fast inversion scheme for the inverse problems. We will study the inverse problem of DOT and FDOT, and discuss an inversion scheme based on the explicit expression of the peak time in the forthcoming paper.

We remark that the expressions of peak time in this paper are obtained under some assumptions for the physical parameters and the object. We assume the size of fluorescence target is very small such that it can be approximated by the Dirac $\delta$-function. However, we can still provide the prior information for general FDOT from the expression of peak time by assuming a point target.

\begin{acknowledgments}
The first author was supported by JSPS KAKENHI (Grant No. JP20F20327). The third author was supported by JSPS KAKENHI (Grant No. JP19K03554). The fourth author was supported by JSPS KAKENHI (Grant No. JP19K04421). The last author was supported by National Natural Science Foundation of China (Grant No. 11971104) and by Natural Science Foundation of Jiangsu Province, China (Grant No. BK20210268).
\end{acknowledgments}

\section*{AUTHOR DECLARATIONS}

The author have no conflicts to declare.

\section*{Data Availability Statement}
Data sharing is not applicable to this article as no new data were created or analyzed in this study.

\bibliography{aipsamp}

\end{document}